\documentclass[twocolumn,showpacs,preprintnumbers,amsmath,amssymb]{revtex4}
\usepackage{tabularx,graphicx}

\usepackage{color}
\usepackage{hyperref}
\hypersetup{
    colorlinks=true,
    linkcolor=blue,
    filecolor=blue,      
    urlcolor=blue,
}

\begin{document}

\newcommand{\beq}{\begin{equation}}
\newcommand{\eeq}{\end{equation}}
\newcommand{\beqn}{\begin{eqnarray}}
\newcommand{\eeqn}{\end{eqnarray}}
\newcommand{\bmath}{\begin{subequations}}
\newcommand{\emath}{\end{subequations}}
\newcommand{\bra}[1]{\langle #1|}
\newcommand{\ket}[1]{|#1\rangle}

\title{On the dynamics  of the Meissner effect}
\author{J. E. Hirsch }
\address{Department of Physics, University of California, San Diego,
La Jolla, CA 92093-0319}

\begin{abstract} 
The question of how a metal becoming superconducting expels a magnetic field is addressed.
It is argued that the conventional theory of superconductivity has not answered this question despite its obvious importance. 
We argue that the growth of the superconducting into the normal region and associated expulsion of magnetic field from the
superconducting region can only be
understood if it is accompanied by {\it motion of charge}   from the superconducting into the normal region. From a microscopic point of view 
it is shown  that the perfect diamagnetism of superconductors requires that superconducting electrons reside in orbits of 
spatial extent $2\lambda_L$, with $\lambda_L$ the London penetration depth. Associated with this physics,
  the spin-orbit interaction of the electron magnetic moment and
the positively charged ionic background gives rise to a ``Spin Meissner'' effect, the generation of a macroscopic spin current near the surface of
superconductors. We point out that both the Meissner and the Spin Meissner effect can be understood dynamically under the 
assumption that the superfluid condensate wavefunction $\Psi(\vec{r})$ does not screen itself, just like the $\Psi(\vec{r})$ for an electron in a
hydrogen atom. 
We argue that the conventional theory of 
superconductivity cannot explain the Meissner effect because it does not contain the physical elements discussed here. \end{abstract}
\pacs{}
\maketitle 

\section{introduction}
It is generally \cite{bcs50} believed that the conventional theory of superconductivity explains the Meissner effect  \cite{tinkham}, the telltale
property of superconductors.
This erroneous belief arises from two related widely held  misconceptions. The first misconception  is that the Meissner effect is an 
``equilibrium phenomenon'' \cite{rick}, that results from a particular form for the relation between current density and magnetic vector
potential in the equilibrium state of a superconductor \cite{klein,bard,bcs}. The second misconception
 is that for a theory of superconductivity
to predict the Meissner effect it is sufficient that it yields a  lower free energy for the system when the magnetic field is excluded than
when the magnetic field remains inside the material \cite{deg}.

The second misconception is easily dispelled  by a counterexample. A superconducting ring with a finite current that generates an
enclosed  quantized magnetic
flux has a higher free energy than the ring with no current and no magnetic flux. Yet there is no mechanism for the system to reach this
lower energy state, and it will remain in the finite current ``metastable'' state forever.

The first misconception is also easily disproved. The Meissner effect is the $process$ by which a metal cooled below its critical temperature in the
presence of a magnetic field in its interior expels the magnetic field and reaches the equilibrium superconductive state with the magnetic field
excluded. A calculation that deals only with quantities in the final state of this process, such as performed in Refs. \cite{rick,klein,bard,bcs} and many
others cannot (by definition) prove that the system will reach this final state nor explain how the system does this. In the absence of 
a description of the process by which the equilibrium state is reached one can equally well  conclude instead that the theory predicts that the system
will forever remain in the initial ``metastable'' state, just like the ring discussed above.

Thus we argue that the current theory of superconductivity does not explain nor predict the Meissner effect. Rather, the most natural conclusion
one should infer 
from the theory is that a
magnetic field either in the interior of a ring, a hollow sphere or a solid body  will remain unchanged
(except perhaps for tiny adjustments to account for quantized flux) when the system is cooled below its critical temperature. 
That this should happen for all three cases was ``proven'' theoretically by Lippmann \cite{lipp}
in 1919  based on Faraday's law. It was confirmed experimentally by Onnes and Tuyn \cite{tuyn} in 1924 for the case of a
hollow sphere. But it was disproved experimentally by Meissner and Ochsenfeld in 1933 \cite{meissner} for the case of a solid body.  We
argue that how superconductors are able to prove Lippmann wrong and ignore Faraday's law   is not explained by the
conventional theory of superconductivity.

There has been a recent claim in the literature that the Meissner effect needs no explanation because any perfect
conductor will show a Meissner effect \cite{essen}. echoeing  a similar claim made earlier \cite{edwards}. These claims are incorrect, as has been proven by
several authors recently \cite{gul,yos} as well as earlier \cite{putterman,taylor}. 
For example, Ref.   \cite{yos} 
 shows that for the Meissner state  to be energetically favorable requires the lowering of energy achieved by
the superconducting condensation energy, which would not exist for a perfect conductor. While this certainly proves 
Ref. \cite{essen}'s claim wrong it does not prove that the Meissner effect is explained by the conventional  theory for the reasons given  above.

Why is the question whether or not the conventional theory of superconductivity explains the Meissner effect relevant? 
Because if the conventional theory cannot explain the Meissner effect it cannot be the correct theory of superconductivity, despite its many apparent successes.
We argue that the  conventional theory   lacks essential physical ingredients that are necessary to explain the Meissner effect.

After discussing in more detail the absence of an explanation
 of the Meissner effect in the conventional theory in the next section, in the 
remainder of this paper we discuss what is required of a theory of superconductivity that can explain the Meissner effect.
The theory of hole superconductivity contains those physical elements \cite{holesc}.

\section{conventional understanding of the Meissner effect}

The London equation, that provides a phenomenological description of the Meissner effect, is usually made plausible 
\cite{am, abr, schrieff} by starting 
from the `acceleration equation' for a perfectly conducting fluid
\beq
\frac{\partial \vec{v}_s}{\partial t}=\frac{q}{m}\vec{E}
\eeq
with $v_s$ the superfluid velocity for carriers of mass $m$ and charge $q$ and $\vec{E}$ the electric field, using Faraday's law
\beq
\frac{\partial}{\partial t}  (\vec{\nabla}\times\vec{v}_s)=-\frac{q}{m c}\frac{\partial \vec{B}}{\partial t} ,
\eeq
integrating Eq. (2) in time and setting the integration constant equal to zero to yield the London equation
\beq
\vec{\nabla}\times\vec{v}_s=-\frac{q}{m c}\vec{B}
\eeq
with $\vec{B}$ the magnetic field. Combined with Maxwell's equation $\vec{\nabla}\times\vec{B}=(4\pi/c)\vec{J}$, with  $\vec{J}=n_s q\vec{v}_s$
the current density, Eq. (3)  yields
\beq
\nabla^2\vec{B}=\frac{4\pi n_s q^2}{m c^2}\vec{B}\equiv \frac{1}{\lambda_L^2}\vec{B}
\eeq
describing the fact that a magnetic field cannot penetrate a superconductor beyond a distance $\lambda_L$ from the surface.

However, integrating Eq. (2) in time yields instead of Eq. (3)
\beq
\vec{\nabla}\times\vec{v}_s(\vec{r},t)-\vec{\nabla}\times\vec{v}_s(\vec{r},t=0)=-\frac{q}{m c}(\vec{B}(\vec{r},t)-\vec{B}(\vec{r},t=0)) .
\eeq
If a normal metal is cooled into the superconducting state in the presence of a spatially uniform 
magnetic field $\vec{B}(\vec{r},t=0)=\vec{B}_0$ throughout its interior,
the initial superfluid velocity $v_s(\vec{r},t=0)=0$ and Eq. (5) yields
\beq
\vec{\nabla}\times\vec{v}_s(\vec{r},t)=-\frac{q}{m c}(\vec{B}(\vec{r},t)-\vec{B}_0)
\eeq
which is $not$ the same as Eq. (3). Quite the contrary, Eq. (6) implies that $\vec{v}_s(\vec{r},t)=0$ and $\vec{B}(\vec{r},t)=\vec{B}_0$ for all times
$t>0$, so that the magnetic field   remains unchanged inside the superconductor. Thus, this derivation certainly does not
describe a $process$ by which the system will achieve the Meissner state described by Eq. (3),  rather it predicts that such a  state will
never be achieved.

Within the conventional (BCS) theory  of superconductivity the Meissner effect is `proven' \cite{klein,bard,bcs} by calculating the London Kernel $K(\vec{q})$ relating the
Fourier components of the current density $\vec{J}(\vec{q})$ and a static magnetic vector potential  $\vec{A}(\vec{q})$ 
\bmath
\beq
\vec{J}(\vec{q})=-\frac{c}{4\pi}K(\vec{q})\vec{A}(\vec{q})
\eeq
and showing that 
\beq 
K(q\rightarrow 0)=\frac{1}{\lambda_L^2}\neq 0
\eeq
\emath
 when the system is described by the BCS wavefunction. Eq. (7) is equivalent to Eq. (3). A great deal of literature
was generated around the question whether Eq. (7)  could be proven in a gauge-invariant fashion, and eventually this was achieved to
everyone's satisfaction  \cite{gauge}. However as discussed in the introduction this does not address the key question which is, how does the system achieve
the BCS state that satisfies Eq. (7)  starting from the initial state that doesn't? 

Furthermore, there is an inherent problem in the linear response argument Eq. (7). It implicitly assumes that the system is initially in the
BCS state, then a uniform magnetic field is created in its interior, and subsequently the system responds to it by generating the Meissner current
that cancels the interior magnetic field. However it is physically impossible to `create' a uniform magnetic field inside a material without 
magnetic lines `cutting through' the material, since Maxwell's equations do not allow magnetic field lines to be created out of nowhere,
they are always necessarily closed. If the system is initially in the BCS state it is a perfect conductor and magnetic field lines cannot cut through
it so the linear response situation cannot be set up. In addition, a system that is already in the BCS state cannot have a magnetic field in its
interior because this is incompatible with global phase coherence. What actually
happens in the Meissner effect is that the system in the normal state is cooled below $T_c$, it is
not in the BCS state initially, and in the process of entering   the BCS state and establishing macroscopic phase coherence the magnetic field is expelled.
This complex process is $not$ described by Eq. (7).

A rationale for the conventional argument seems to be \cite{scal} that when the system is cooled below $T_c$ in the presence of a weak uniform
magnetic field, it is expected that thermal fluctuations will lead to the transition into the superconducting state with the field expelled 
because it  has lower free energy
than the normal state. However, we argue that thermal fluctuations are local and such random fluctuations cannot generate the global surface current
necessary to expel the magnetic field except with vanishing probability for a macroscopic system.

A somewhat more general   `proof' of the Meissner effect \cite{feyn}  starts from the assumption that the many-electron superfluid condensate  of a superconductor can be described by a complex macroscopic wavefunction $\Psi(\vec{r})$. 
This was first done in a phenomenological way by Ginzburg and Landau \cite{gl}, later it was shown that such a description can be derived from
BCS theory  under certain conditions  \cite{gorkov,scully}, and the predictions of Josephson  \cite{joseph} and subsequent experimental
verifications  \cite{merc}  established without doubt that it is a correct description of
the superfluid condensate that captures the essence of superconductivity, whether BCS theory is valid or not.
The amplitude of this macroscopic  wave function $\Psi(\vec{r})$ is related to the density of superconducting carriers $n_s$
\beq
\Psi (\vec{r})=|\Psi(\vec{r})|e^{i\phi(\vec{r})}=n_s^{1/2}e^{i\phi(\vec{r})}
\eeq
and the gradient of the phase $\phi(\vec{r})$ is related to the superfluid velocity $\vec{v}_s$ according to the relation 
\beq
\vec{v}_s=\frac{\hbar}{m}   \vec{\nabla} \phi-\frac{q}{mc} \vec{A} .
\eeq
The wavefunction $\Psi(\vec{r})$ describes the state
of all the Cooper pairs in the system, that share a common phase   $\phi(\vec{r})$ that  is coherent over macroscopic distances \cite{merc}. 
Many of the unique  and universal properties of superconductors such as   flux quantization and the variety of phenomena exhibited by
Josephson junctions and `weak links' follow from this simple macroscopic description.
Upon taking the curl on both sides of Eq. (9) the London Eq. (3) results, hence this is assumed to be a
proof of the Meissner effect. However once again, in assuming that the system is described by the macroscopic 
wavefunction Eq. (8) one is assuming that the magnetic field has already been expelled, since Eq. (8) is not valid in the
presence of a magnetic field in the interior of the superconductor, and the $process$ by which the system
reaches or doesn't reach this state starting from the initial state with the magnetic field in the interior is not
discussed.

Finally, an `energetic' argument for the Meissner effect is given for example in ref. \cite{deg}, where it is shown that the magnetic field distribution
where the magnetic field is excluded except in a region within $\lambda_L$ of the surface minimizes the free energy. 
However as already emphasized, this does not explain the process of field expulsion nor predicts that the state will be reached.

In summary, we argue that these arguments, which are generally assumed to prove that the conventional theory of superconductivity
describes the Meissner effect, in fact do not do so, and leave completely open the question whether or not the conventional theory 
of superconductivity can describe the
Meissner effect.

More recently, there have been calculations describing the normal-superconductor transition in a magnetic field \cite{dorsey, goldenfeld} using 
time-dependent Ginzburg Landau theory \cite{tdgl1,tdgl2}. In TDGL theory it is $assumed$ that the superconducting
order parameter ($\Psi(\vec{r})$) relaxes exponentially to its equilibrium value in a non-equilibrium situation.
However, no justification for this assumption is presented. In fact, very recent work claims \cite{bcstdgl1,bcstdgl2} that this assumption of TDGL theory
is incorrect and that  within BCS-Bogoliubov-de-Gennes theory the superconducting order parameter will $not$ relax
spontaneously to its equilibrium value.

 \section{how a perfect conductor expels a magnetic field}
 Let us consider again the equation of motion for a perfectly conducting fluid. Eq. (1) is not quite correct because it ignores  the difference between
total and partial time derivatives, and because the magnetic Lorentz force is omitted from the right side. Taking both of these facts into account
leads instead of to Eq. (2) to the equation \cite{londonbook}
\bmath
\beq
\frac{\partial \vec{w}}{\partial t}= \vec{\nabla}\times   (\vec{v_s} \times\vec{w})
\eeq
with
\beq
\vec{w}=\vec{\nabla}\times\vec{v_s}+\frac{q}{m c}\vec{B}
\eeq
\emath
the `generalized vorticity'.  The London Eq. (3) is 
\beq
\vec{w}(\vec{r},t)=0
\eeq
and the initial condition at the moment a system is cooled into the superconducting state in a uniform magnetic field $B_0$  is
\beq
\vec{w}(\vec{r},t=0)=\vec{w}_0=\frac{q}{mc}\vec{B}_0  .
\eeq
To explain the Meissner effect along these lines
one has to explain how $\vec{w}$ evolves in time from its initial value Eq. (12) to its final value Eq. (11) 
at all $\vec{r}$'s at a later time.

Let us assume for simplicity a cylindrical geometry with $\vec{B}$ in the $z$ direction. $\vec{w}=w\hat{z}$ and Eq. (10a) can be rewritten as
\beq
\frac{\partial  w}{\partial t}= -\frac{1}{r}  \frac{\partial}{\partial r}(r wv_r)
\eeq
which indicates that a {\it radial velocity} $v_r\neq 0$ of the fluid is a prerequisite for $w$ to change in time. 
 In the absence of
a radial velocity, $\partial w/\partial t=0$ and $w(\vec{r},t)=w_0$ for all times and the magnetic field in the interior of the metal becoming superconducting
would remain unchanged, in agreement with
 Lippmann \cite{lipp} and in contradiction with experiment.

Eq. (13) can be rewritten as
\beq
\frac{\partial  w}{\partial t}=-\vec{\nabla}\cdot (w\vec{v_s})
\eeq
which is a continuity equation. It says that for $w$ to change locally it has to be carried away  by the fluid flow. Integrating Eq. (14) over a cylinder of radius R and using Gauss' theorem yields
\beq
\int_0^R dr  r          \frac{\partial}{\partial t}        w(r,t)=-Rw(R,t)v_r(R,t)
\eeq
and integrating over time and assuming at time T the Meissner state $w(r,T)=0$ for all $r\leq R$ has been reached yields the condition
\beq
w_0 =\frac{2}{R} \int_0^Tdtw(R,t)v_r(R,t)    .
\eeq
According to this calculation in order to achieve the Meissner state in the interior of a cylindrical superconductor
of radius $R$,  i.e. $w(r)=0$ for all $r\leq R$, the entire superfluid has to flow $out$ of the material  carrying the generalized vorticity $w$ with it. 

Of course superconductors don't do that. The reason this calculation does not apply is that it assumes the system first becomes a perfect conductor
and subsequently expels the magnetic field As we discuss in the next section, this is not what occurs in reality.
Nevertheless this analysis is  useful because it indicates  that an $outflow$ of charge is necessary to understand the
dynamics of the Meissner effect.

 \section{How  the transition occurs}

In a recent paper \cite{pipp} we have discussed   the kinetics of the normal-superconductor transition in  a magnetic field. 
Figure 1 shows three  generic scenarios. As discussed in \cite{pipp}, the simplest  scenario (a), where a current develops near the surface
of the sample and the magnetic field is uniformly expelled can be excluded on theoretical as well as experimental grounds. 
In scenario (b) superconducting regions nucleate at random and expand. Scenarios (b) and (c) are in essence similar so we will limit ourselves to scenario (c), where the superconducting
phase expands from a nucleus at the center towards the surface with a surface current that excludes the magnetic field in the interior of the superconducting
region. This has been implemented experimentally in a cylinder by applying a slightly smaller magnetic field in the central region \cite{faber}.

Both in scenarios (b) and (c) a Faraday electric field $E$ exists at and near  the boundary
  of the domains because the magnetic field is changing as the domains
  expand. The direction of this electric field is opposite to the
  direction of the boundary current as shown in Fig. 1(c), to induce a current opposing the change in magnetic flux
  (Lenz's law). Such a
  counterclockwise electric 
  field also exists throughout the interior of the sample in (a) during the
  transition.
  
   \begin{figure}
 \resizebox{8.5cm}{!}{\includegraphics[width=6cm]{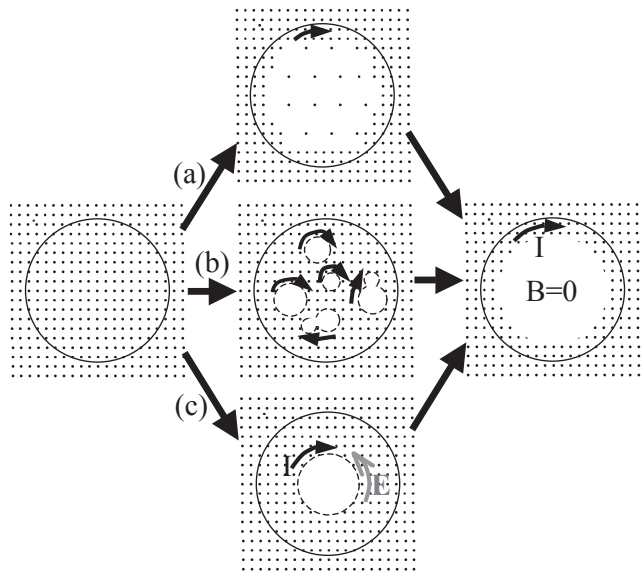}}
 \caption { Three conceivable  routes for the magnetic field expulsion in a cylindrical superconductor. 
 The dots represent magnetic field lines coming out of the paper. The arrows give the
 direction of the currents ($I$).  In (a),
 a current starts developing within $\lambda_L$ of the surface of the cylinder that gradually
 increases in magnitude, causing a gradual uniform decrease of the magnetic field in the interior.
  In (b), domains of loop currents start at various random locations  that nullify the magnetic fields in 
 their interior, that gradually expand their
 radii and coalesce and merge with neighboring domains. In (c), a single circular domain starts at
 the center and expands in radius until it reaches the boundary of the cylinder.
   }
 \label{figure1}
 \end{figure}

Focusing on scenario (c), an explanation of the Meissner effect has to explain how this current at the surface of the expanding superconducting
region is generated in the first place, how it is maintained on the face of the Faraday electric field that tries to
suppress it, how its momentum and kinetic energy is transferred radially outward as the superconducting
phase expands, and how this growing momentum (and angular momentum) is compensated so as to not violate 
fundamental conservation laws. None of these questions is addressed by theoretical treatments of this process
based on the  conventional BCS-London-TDGL
theory of superconductivity \cite{dorsey,goldenfeld}.

   \begin{figure}
 \resizebox{8.5cm}{!}{\includegraphics[width=6cm]{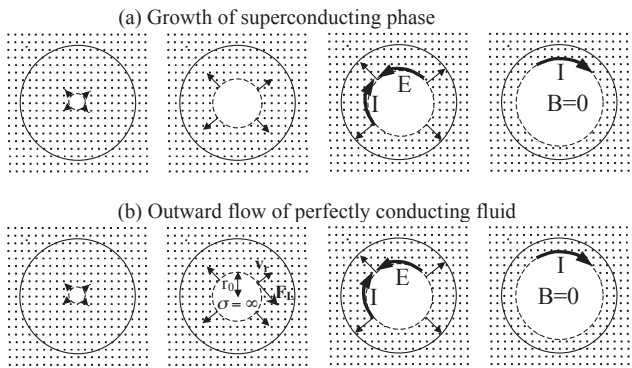}}
\caption { (a) More stages of the process of magnetic field expulsion as a system goes superconducting according to  the scenario (c) of Fig. 1. 
 At and near the boundary of the expanding field-free region there is an electric field pointing counterclockwise, due to the
 changing magnetic flux and Faraday's law, and a current flowing clockwise that cancels the applied magnetic field
 in the interior.  
 (b) Schematic depiction of a perfectly conducting fluid ($\sigma=\infty$)  that flows radially
 outward with radial velocity $v_r$. The carriers at the boundary experience a Lorentz force.
 Assuming the sign of the charge $q$ is positive for definiteness the
 Lorentz force $F_L=(q/c)v_rB$ points in the clockwise direction. The resulting
 electric current $I$ at the boundary flows clockwise (for negative charge carriers the
 Lorentz force would be in opposite direction, the current in the same direction), generating a magnetic
 field opposite to the external field so that no magnetic field penetrates the perfect 
 conductor.   }
 \label{figure1}
 \end{figure}

 Figure 2 (a)  shows more stages of the process of magnetic field expulsion (c) of Fig. 1.
BCS-London-TDGL theory would say that the physics driving the expansion of the
field-free region is the superconducting condensation energy.
 The energy of the system is lowered as Cooper pairs form and become phase coherent, condensing
 into the macroscopic superconducting state described by $\Psi(\vec{r})$. Because the
 establishment of phase coherence requires that no magnetic field exists in that region, the region has to 
 become field free, 
 and this occurs through a BCS order parameter
 spontaneously growing and relaxing towards its equilibrium value \cite{dorsey,goldenfeld}. 
 This is a phenomenological treatment that   does not explain  how
the condensation into the BCS state  is connected to the azimuthal motion of the carriers
that needs to be generated at the boundary, and in particular what is the force that drives these carriers to move in direction opposite to
the electric force exerted on them by the Faraday electric field. Within BCS-London-TDGL theory there is no $radial$ charge motion associated with the outward
motion of the phase boundary between superconducting and normal regions.

The behavior shown in Fig. 2 (a) would also occur if a core of high density perfectly conducting material 
(infinite conductivity) would expand radially outward, as shown schematically in Fig. 2 (b). 
Here the charge carriers are moving radially out and   there is  a Lorentz force $F_L$ acting on them giving them an 
azimuthal velocity (in the clockwise direction assuming positive carriers) such that they generate a magnetic field opposite to the applied one,
thus not allowing the magnetic field to penetrate the interior of the perfectly conducting region (except to within a distance $\lambda_L$ of its surface). 
We discuss the process  quantitatively in what follows in  a planar geometry for simplicity.

\section{meissner effect in planar geometry}
Figure 3 shows the processes of Fig. 2 in a planar geometry. The magnetic field points in the $z$ direction, the phase boundary
advances in the $+x$ direction, and the Faraday electric field and Meissner current point in the $y$ direction. 
$x_0(t)$ is the boundary of the superconducting region, moving upward at rate $dx_0/dt$. The magnetic and
electric fields for $x\leq x_0$ are given by 
\bmath
\beq
B(x)=B_0e^{(x-x_0)/\lambda_L}
\eeq
\beq
E_y(x)=\frac{B_0}{c} \frac{dx_0}{dt}e^{(x-x_0)/\lambda_L}
\eeq
with $B_0\hat{z} $ the applied magnetic  field and $E_y$ in the positive $y$ direction. 
Assuming current carriers of charge $q$, mass $m$ and density $n_s$, their   speed in the $y$ direction is 
\beq
v_y(x)=-\frac{c}{4\pi n_s q \lambda_L}B_0e^{(x-x_0)/\lambda_L}
\eeq
\emath
parallel (antiparallel) to the current, which flows in the $-\hat{y}$ direction, for  $q>0$ ($q<0$).  For $x>x_0$ we assume for simplicity
\bmath
\beq
B(x)=B_0
\eeq
\beq
E_y(x)=\frac{B_0}{c} \frac{dx_0}{dt}
\eeq
\emath
This assumption corresponds to the treatment of ref. \cite{pipp} for the particular case where the magnetic field that is being expelled
is very close to the critical field ($p\rightarrow 0$ in the notation of ref. \cite{pipp}).

\begin{figure}
 \resizebox{8.5cm}{!}{\includegraphics[width=6cm]{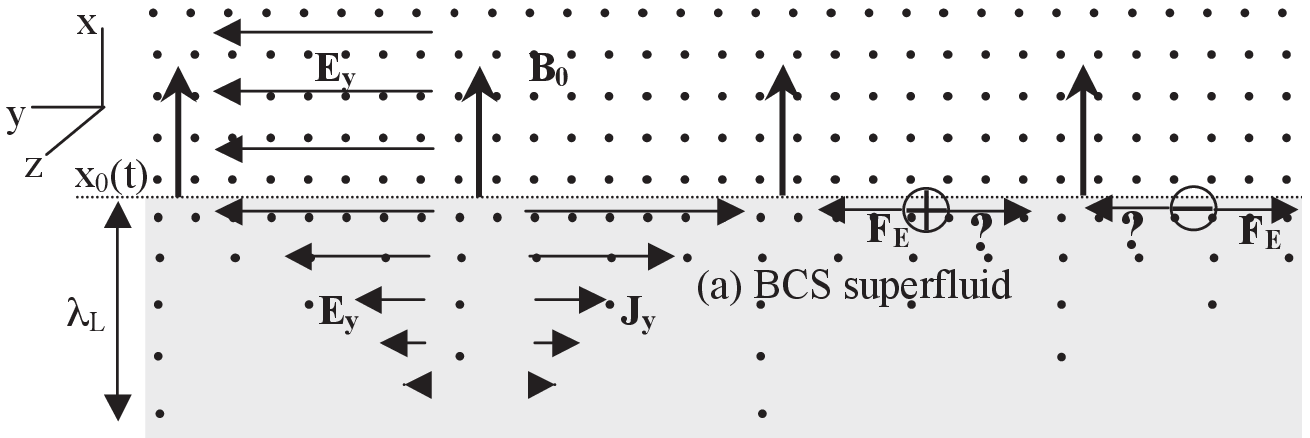}} 
 \newline
  \newline
   \newline
 \resizebox{8.5cm}{!}{\includegraphics[width=6cm]{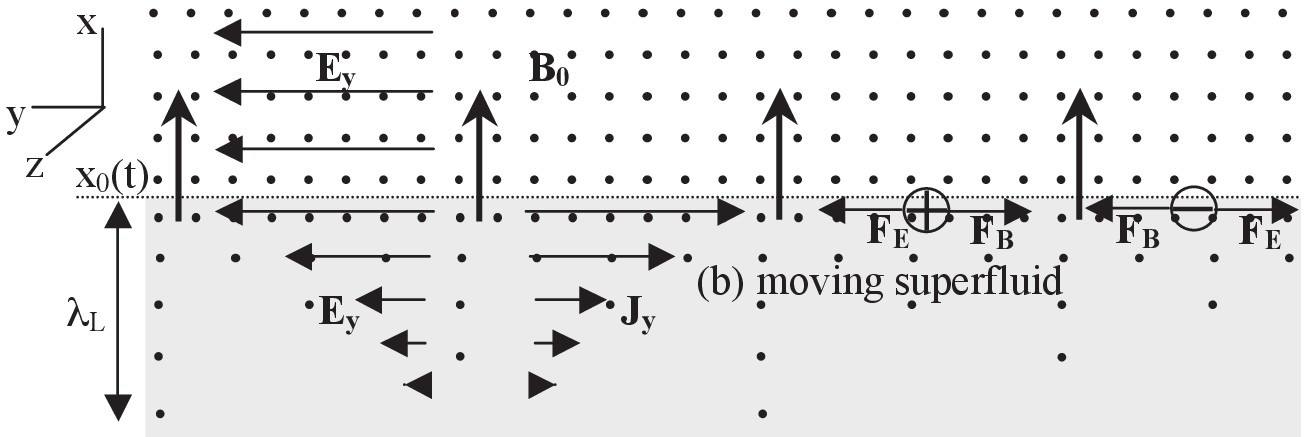}}
 \caption { Magnetic field $B_0$ points out of the paper. (a) Superconducting-normal phase boundary $x_0(t)$ moving in the $+x$ direction with no associated charge flow in the $x$ direction.   (b)  Superfluid moving in the $+x$ direction with boundary at $x_0(t)$. For both cases the magnetic field
 is expelled from the region $x<x_0(t)$ by the current $J_y$ flowing within $\lambda_L$ of the boundary, and a Faraday field $E_y$ exists due to the changing
 magnetic field.  The electric Lorentz force ($F_E$) drives the carriers in direction $opposite$ to their direction of motion. Only for case (b) the carriers
 also experience a  magnetic Lorentz force $F_B$ driving them in the direction of their motion, opposite to  $F_E$ and of equal magnitude. 
}
 \label{figure1}
 \end{figure}

Equations  (17a,b,c)  follow from the relation between current and velocity
$\vec{J}=n_sq\vec{v}$, the expression for the London penetration depth Eq. (4), Maxwell's equations
\bmath
\beq
\vec{\nabla}\times\vec{B}=\frac{4\pi}{c} \vec{J}==>\frac{\partial B}{\partial x}=-\frac{4\pi}{c}J_y=-\frac{4\pi n_s q}{c}v_y
\eeq
\beq
\vec{\nabla}\times\vec{E}=-\frac{1}{c} \frac{\partial\vec{B}}{{\partial t}}==>\frac{\partial E_y}{\partial x}=-\frac{1}{c}\frac{\partial B}{\partial t}
\eeq
and London's equation
\beq
\vec{\nabla}\times\vec{J}=-\frac{n_s q^2}{mc}\vec{B} ==> \frac{\partial v_y}{\partial x}=-\frac{q}{mc}B  .
\eeq
\emath
The electromagnetic force acting on the carriers is
\beq
m\frac{d\vec{v}}{dt}=\frac{q}{m}\vec{E}+\frac{q}{mc}\vec{v}\times\vec{B}
\eeq
and its component in the $y$ direction yields
\beq
\frac{\partial v_y}{\partial t}+v_x\frac{\partial v_y}{\partial x}=\frac{q}{m}E_y-\frac{q}{mc}v_xB
\eeq

For the case of Fig. 3(b) (moving superfluid) we have
\beq
v_x=\frac{dx_0}{dt}
\eeq
and the terms on the right-hand side of Eq. (21) satisfy
\beq
F_E=\frac{q}{m}E_y=\frac{q}{mc}v_x B=F_B
\eeq
so that the electric and magnetic forces in Eq. (21) exactly cancel out. The left-hand side of Eq. (21) is also identically zero from
Eqs. (17c) and (22). 
Carriers are accelerated by the magnetic Lorentz force and decelerated by the electric  force arising from Faraday's
field, which exactly cancel in steady state. Thus, a dynamical explanation of flux expulsion is provided by these equations: {\it the
magnetic Lorentz force acting 
on the outflowing charge drives the current that nullifies the magnetic field in the interior against the Faraday electric force}.

For the processes in Figs. 3 to happen requires a driving force in the $+\hat{x}$ direction. 
The way this works is very clear for case (b), moving superfluid:   there is a magnetic
Lorentz force on the $J_y$ current carriers in the $-\hat{x}$ direction:
\beq
F_x=\frac{q}{c}(\vec{v}\times\vec{B})_x=\frac{q}{c}v_y B
\eeq
so that an equal and opposite force needs to be applied in the $+\hat{x}$ direction for the superfluid to advance in the $+\hat{x}$ direction. This
requires expenditure of energy per unit time  per carrier $-F_xv_x$, so that the energy per unit area per unit time spent is,
from Eqs. (17a), (17c) and (24)
\beq
\int_{-\infty}^{x_0} dx n_s (-F_x v_x)=\int_{-\infty}^{x_0} dx   \frac{B(x)^2}{4\pi\lambda_L} v_x=\frac{B_0^2}{8\pi}\frac{dx_0}{dt}
\eeq
which equals the rate of change of magnetic energy per unit area as the boundary moves, as required 
by energy conservation. 
There is no energy dissipated in this process.

The same rate of energy expenditure is required for the process of Figure 3(a) to expel the magnetic field. This energy obviously is supplied by the 
superconducting condensation energy. However  since in this case $v_x=0$, Eq. (25) cannot be used to understand how
the condensation energy causes the phase boundary to advance. The force $F_x$ given by Eq. (24) still exists
in this case, but since carriers don't move in the $x$ direction this force does not do any work. 
The work done by the condensation process in moving the phase boundary in the $+\hat{x}$ direction
without displacing charge carriers in the $+\hat{x}$ direction occurs through some unknown way.

In addition  since charge does not move 
in the direction of the moving phase boundary there is no magnetic Lorentz force driving the Meissner current 
in Fig. 3(a) and 
Eq. (21) is not satisfied. Eq. (21) becomes
\beq
\frac{\partial v_y}{\partial t} =\frac{q}{m}E_y 
\eeq
which certainly does not describe Fig. 3(a) since it says that  the current should be flowing in the opposite direction, parallel 
rather than antiparallel to $E_y$.
In steady state the left side in Eq. (26) is zero and there is an uncompensated right-hand-side.

One would have to include other dynamical effects to `fix' Eq. (21) and to give a dynamical explanation of the energy
transfer process for the case of Fig. 3(a) other than Eq. (25). This has not been   done within the conventional theory.
We argue that the dynamics described by Eqs. (20) and (25), requiring charge motion in the same direction
as the phase boundary motion,  is the only physical  way to describe the flux expulsion.

\section{backflow}
As discussed in the previous section, the strong similarity in the physics shown in Figs. 3(a) and 3(b) is compelling evidence that in the process of a metal becoming
superconducting an $outflow$ of charge in direction normal to the superconductor-normal metal phase boundary into the
normal region takes place. This explains both how the carriers in the supercurrent flowing parallel to the phase boundary
can move in direction opposite to the Faraday electric force, and how the
superconducting condensation energy is used to do the work required to expel the magnetic field: the condensation results in an $electromotive$
 $force$ \cite{emf} driving
the carriers to move in the $+\hat{x}$ direction, i.e. from the superconducting into the normal region, opposing the electromagnetic force Eq. 24.

In order for this not to create an enormous charge imbalance it requires a backflow of normal carriers in the $-\hat{x}$ direction. 
Figure 4 shows schematically how this can happen. The layer of superfluid of thickness $\lambda_L$ next to the phase boundary
with the velocity pattern given by Eq. (17c) moves forward a distance $\lambda_L$, and an equal amount of normal fluid moves
backward, and in the process becomes superfluid. This satisfies energy and momentum conservation and gives a simple description of the
process. The driving force for the backflow is the electric field in direction perpendicular to the phase boundary created by the forward flow.

\begin{figure}
 \resizebox{8.5cm}{!}{\includegraphics[width=6cm]{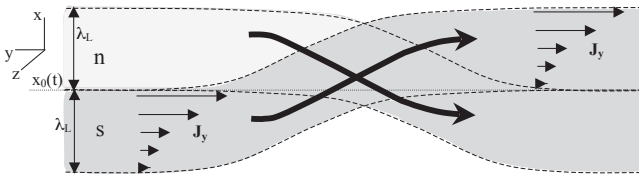}} 
 \caption { Schematic depiction of how the superfluid-normal phase boundary moves into the normal region. Superfluid layer of thickness
 $\lambda_L$ carrying the screening
 current $\vec{J}_y$ moves forward and an equal amount of normal fluid moves backward and becomes superconducting.
 }
 \label{figure1}
 \end{figure}

\begin{figure}
 \resizebox{8.5cm}{!}{\includegraphics[width=6cm]{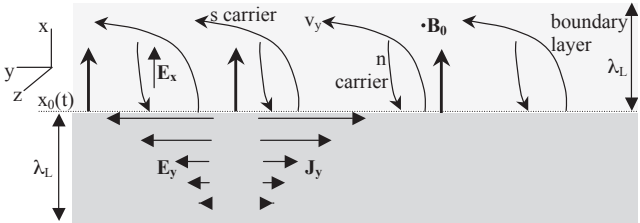}} 
 \caption { As carriers become superconducting (s carriers) they thrust forward into the normal region over a boundary layer of thickness $\lambda_L$, and
 are deflected by the Lorentz force acquiring speed $v_y=-c/(4\pi n_s q \lambda_L)B_0$. This process creates an electric field $E_x$ in the $-\hat{x}$ direction that drives normal carrier (n carrier) backflow. Here, ``n'' stands both for ``normal'' and ``negative''. The normal carriers do not acquire a large $v_y$ in opposite direction because they scatter off
 impurities and transfer their $y-$momentum to the lattice.
 }
 \label{figure1}
 \end{figure}    
 
 However, considered in detail the process is likely to be more complicated. We propose that the forward motion occurs for carriers that are
 {\it in the process} of becoming superconducting, which are initially (essentially) not moving in the $y$ direction and acquire a large
 velocity parallel to the phase boundary through the magnetic Lorentz force. 
As discussed in Ref. \cite{pipp}, in order for a carrier moving in the $+\hat{x}$ direction to acquire the transverse velocity Eq. (17c) through the
action of the magnetic Lorentz force requires a displacement over a distance $\lambda_L$ in the $x$ direction. 
Thus, we envision a process of forward flow and backflow over a boundary layer of thickness $\lambda_L$   as the phase boundary
advances, as shown schematically in Fig. 5. As carriers become superconducting they thrust forward a distance $\lambda_L$, which creates an electric field in the
$+\hat{x}$ direction (assuming these are negatively charged carriers) causing backflow of normal carriers in this layer. As the phase boundary advances at rate $v_0=dx_0/dt$, normal (negatively charged) carriers in the
boundary layer are back-flowing at speed $v_0$, or equivalently $positive$ normal carriers (holes) move forward together with the phase boundary.
Because normal
carriers scatter off the lattice they do not acquire a large $v_y$ from the action of the magnetic Lorentz force, instead they transfer their
$y-$momentum to the lattice as a whole, thus accounting for momentum conservation. 

This gives a phenomenological description of the magnetic field expulsion process that resolves the puzzles of the Meissner effect 
regarding forces and momentum conservation. A complete microscopic description has not yet
been proposed. A semiclassical explanation based on the proposal that superconducting carriers reside in mesoscopic orbits of
radius $2\lambda_L$ \cite{sm} is discussed in the following sections.

\begin{figure}
 \resizebox{8.5cm}{!}{\includegraphics[width=6cm]{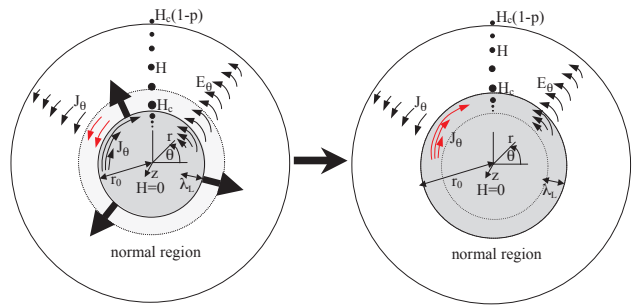}} 
 \caption { Expansion of superconducting phase in a cylindrical geometry. On the left, the dark grey circle is superconducting and the light 
 grey ring (of thickness $\sim \lambda_L$) is about to become superconducting. As the ring becomes superconducting the current direction in the ring switches from  the $+\hat{\theta}$ to the $-\hat{\theta}$ direction (red arrows) which is opposite to the electric field direction.
 }
 \label{figure1}
 \end{figure}

  For completeness we show in figure 6   the current and fields in a cylindrical geometry. 
  The magnetic field that is being expelled is $H_c(1-p)$ with $p>0$, and it 
  increases to $H_c$ at the phase boundary due to the magnetic field generated by the eddy current in the normal region.
  The equations are given in Ref. \cite{pipp}. Again, the process is simply understood assuming the outward expansion of
  the phase boundary is accompanied by radial outflow of charge, driven by a radial electromotive force that pushes the charges against
  a radially inward pointing magnetic Lorentz force. An inward backflow of normal current (not shown) compensates for the charge 
  imbalance. In the absence of radial outflow of charge there is no mechanism to explain how the current reverses its
  azimuthal direction (red arrows) and flows against the direction of the electric field when the phase boundary advances, nor to explain
  how the growing angular momentum of the Meissner current is compensated \cite{pipp,missing}.

  \section{Meissner effect and $2\lambda_L$ orbits}
  
  Consider a superconducting long cylinder in an applied magnetic field $H$. The Meissner current that nullifies the magnetic field in the interior 
  flows within a distance $\lambda_L$ of the surface with speed
  \beq
  v_s=-\frac{e \lambda_L}{m_e c} H .
  \eeq
  with $e$ and $m_e$ the electron charge and mass.
  This is easily seen from the relation between magnetic field and vector potential for the cylinder, $A=H\lambda_L$. 
  The Meissner current density is
  \beq
  j_s=n_s e v_s=\frac{n_s e^2}{m_e c}\lambda_L H
  \eeq
  and from Maxwell's equation $\vec{\nabla} \times \vec{H}=(4\pi/c)\vec{j}_s$ it follows that
  \beq
  H=\frac{4\pi}{c} j_s \lambda_L
  \eeq
  and combining Eqs. (28) and (29) yields
  \beq
  \frac{1}{\lambda_L^2}=\frac{4\pi n_s e^2}{m_e c^2}  .
  \eeq
  The magnetization per unit volume is given by 
  \beq
  M=\frac{H}{4\pi}=\frac{j_s}{c}\lambda_L=\frac{en_s}{c}\lambda_L v_s
  \eeq
It is reasonable to assume that
 the magnetization results from a superposition of elementary magnetic moments $\vec{\mu}$ resulting from the orbital motion of
  each superconducting electron:
  \beq
  \vec{M}=n_s\vec{\mu} .
  \eeq
  This yields for the magnetic moments
  \beq
 \mu=\frac{e\lambda_L}{c}v_s
  \eeq
  and from the general principle that the magnetic moment and orbital angular momentum $\vec{\ell}$ of each electron are related by
  \beq
  \vec{\mu}=\frac{e}{2m_e c}\vec{\ell}
  \eeq
  we deduce that the angular momentum associated with each of these electrons is
  \beq
  \ell=\frac{2m_e c}{e}\mu=m_e v_s (2\lambda_L) .
  \eeq
Eq. (35) implies that  the Meissner current that nullifies the magnetic field in the interior of a superconductor results
  from each superfluid carrier moving with speed $v_s$ in a mesoscopic orbit of radius $2\lambda_L$. Just like in a magnetic material,
  the superposition of elementary currents results in a macroscopic surface current.
 
It can be seen  that    the fact that electrons reside in $2\lambda_L$ orbits by itself  is  sufficient to 
  give perfect diamagnetism, as follows.  For an electron in an orbit of radius $r$, application of an external field $H$ yields through Faraday's law a tangential electric field
  \beq
  E=-\frac{r}{2c}\frac{\partial H}{\partial t}
  \eeq
  and from the equation of motion $m_e dv/dt=eE$,
  \beq
  \frac{dv}{dt}=-\frac{er}{2m_ec}\frac{\partial H}{\partial t}
  \eeq
  so that Eq. (27) results upon integration for $r=2\lambda_L$. However there is a subtle question of self-consistency that we discuss in the next section.

   \section{Meissner effect and magnetic susceptibility}
   An external magnetic field $H$ applied to a material gives rise to a magnetization $M=\chi H$ and a total magnetic field
   \beq
   B=H+4\pi M=(1+4\pi \chi)H   .
   \eeq
   A perfect diamagnet does not allow any magnetic field in its interior, hence is defined by $\chi=-1/(4\pi)$ according to Eq. (38).
   This is usually assumed to be the susceptibility of superconductors.
   
   However, Pippard points out \cite{pipchi} that this is a misconception when applied to superconductors because
   ``the mean field responsible for magnetizing an extended unit is not $H$ but $B$'', from which it follows that
   $M=\chi B$ rather than $M=\chi H$  in Eq. (38), hence 
 \beq
   B=\frac{1}{1-4\pi \chi}H ,
   \eeq
   and  ``perfect diamagnetism demands that $\chi$ be infinitely negative'' \cite{pipchi}. The same argument is made by
   Tinkham \cite{tinkchi}.
   
   Consider a system of electrons of density $n_s$ per unit volume in orbits perpendicular to an applied magnetic field $H$. The
   Larmor diamagnetic susceptibility is given by
   \beq
   \chi=-\frac{n_s e^2}{4m_e c^2}<r^2>
   \eeq
   where $\sqrt{<r^2>}\equiv \bar{r}$ gives the spatial extent of the orbit, i.e. its `radius'. For \beq
   \bar{r}=2\lambda_L
   \eeq
   the diamagnetic susceptibility Eq. (40) is $\chi=-1/4\pi$ according to Eq. (30), 
   and we have argued in the past that this should apply to superconductors \cite{missing}. Instead,
   Pippard and Tinkham   argue, as discussed above, that $ \bar{r}=\infty$ is required in superconductors to give
   $\chi=-\infty$ and $B=0$ (Eq. (39)).
   
   Indeed it would appear at first sight that the Pippard-Tinkham point of view is the correct one. The orbits of radius given 
   by Eq. (41) are highly overlapping since 
   \beq
   2\lambda_L=n_s^{-1/3}\sqrt{\frac{n_s^{-1/3}}{\pi r_c}} >>n_s^{-1/3}
   \eeq
 where $r_c=e^2/m_ec^2$ is the classical electron radius, much smaller than 
 the inter-electron distance $n_s^{-1/3}$. Therefore one would expect
 the magnetic field generated by one orbit to reduce the magnetic field affecting nearby overlapping orbits, and the net magnetic
 field resulting from application of an external magnetic field $H$ to be, from Eq. (39)
 \beq
 B=\frac{1}{1-4\pi(-1/4\pi)}H=\frac{H}{2}
 \eeq
  which is not the complete flux expulsion observed in superconductors.
  
  The same result is obtained from the analysis in the previous section. Faraday's law Eq. (36) should involve the total field $B$ rather than $H$,
  hence instead of Eq. (37) we have
      \beq
  \frac{dv}{dt}=-\frac{er}{2m_ec}\frac{\partial B}{\partial t}
  \eeq
  The total magnetic field is weaker than the applied magnetic field due to the counterfield generated by the other electrons in the system:
  \beq
B=H+4\pi M=H+4\pi n_s \mu=H-\frac{2\pi n_se r}{c}v
\eeq
using that $\mu=-erv/2c$ for electrons with speed $v$ in orbits of radius $r$.  Replacing in Eq. (44) and performing the time integration yields
\beq
v=-\frac{\frac{er}{2m_ec}}{1+\frac{\pi n_s e^2}{m_e c^2}r^2}H
\eeq
which for $r=2\lambda_L$ yields
\beq
v=-\frac{e\lambda_L}{2m_ec}H=\frac{v_s}{2} .
\eeq
predicting  that electrons aquire only half of the required speed Eq. (27) to nullify the applied magnetic field $H$, hence that the applied
magnetic field will not be nullified in the interior but only reduced by a factor of 2, in agreement with Eq. (43). 

This result is puzzling because a perfect classical conductor
will perfectly screen an applied magnetic field in its interior. We conclude that to make this semiclassical model agree with the classical behavior
requires that Eq. (37) rather than Eq. (44) applies, in other words that {\it electrons in the superfluid are only affected by the external magnetic  field and not by the
magnetic field generated by other electrons in the superfluid}.

Thus,   to understand the perfect diamagnetism of superconductors in terms of $2\lambda_L$ orbits as suggested by Eqs. (32) and (35) it is necessary to   assume
  that the magnetic field created by  superfluid electrons does  not affect the superfluid itself.   This is not an implausible assumption.
  It  will occur if the superconducting electrons are described by a macroscopic wavefunction
  $\Psi(\vec{r})$ that does not screen itself.  Thus, when an external magnetic field
  is applied all the components of the macroscopic wavefunction  are subject to  the entire applied field rather than to the external field reduced 
by the action of other components of the wavefunction. Under those conditions, Eq. (37) rather than Eq. (44) applies and the induced magnetic field 
for electrons in orbits of radius $2\lambda_L$ is precisely
  of the magnitude needed to completely cancel the interior magnetic field.
  
    This key property of the quantum-mechanical wave function was   recognized from the outset of the development of quantum mechanics.
 Schr\"odinger expressed it clearly in his 1952 paper on the meaning of wave mechanics \cite{schr}: {\it ``The original wave~mechanical model of the hydrogen
atom is not self-consistent. The electronic cloud effectively
shields the nuclear charge towards outside, making up a neutral
whole, but is inefficient inside; in computing its structure
its own field that it will produce must not be taken into account,
only the field of the nucleus.''}  The reason  Schr\"odinger  spelled this out in detail is because it contradicted his intuition and physical expectation that
$|\Psi(\vec{r})|^2$ would represent the $charge$ $density$ of the electron at position $\vec{r}$ rather than the $probability$ of finding the electron at $\vec{r}$.

  As we discuss in the next section, the same physics explains how a magnetic field is expelled from the interior of a system becoming 
  superconducting, and furthermore we argue that this process cannot be explained in the absence of this physics.

      \section{Dynamical explanation of the Meissner effect}
   
   The Landau diamagnetic susceptibility of conduction electrons is given by
   \beq
   \chi=-\frac{1}{3} \mu_B g(\epsilon_F)
   \eeq
   with $g(\epsilon_F)$ the density of states at the Fermi energy and $\mu_B$ the Bohr magneton. Using the expression for the free electron density of states 
   $g(\epsilon_F)=3n/2\epsilon_F$ it is found that the Landau susceptibility Eq. (48) is the Larmor diamagnetic susceptibility 
   Eq. (40) for
   \beq
   <r^2>=k_F^{-2}   .
   \eeq
   and $n=n_s$, hence that in the normal state electrons reside in microscopic non-overlapping orbits of radius $k_F^{-1}$. 
   Therefore, in the process of becoming superconducting and establishing phase coherence electrons {\it expand their orbits} from
   radius $k_F^{-1}$ to radius $2\lambda_L$.

         \begin{figure}
 \resizebox{5.5cm}{!}{\includegraphics[width=6cm]{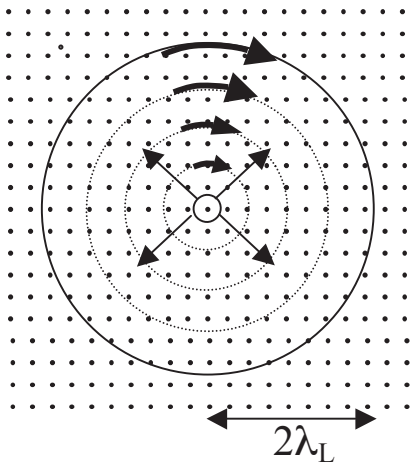}}
 \caption { A single charge carrier of positive charge expanding its orbit in a magnetic field pointing out of the paper acquires an azimuthal velocity in the clockwise
 direction, and itself generates a (small) magnetic field pointing into the paper.   }
 \label{figure1}
 \end{figure}

   The process of expansion of electronic orbits from radius $k_F^{-1}$ to radius $2\lambda_L$ provides a dynamical explanation of the Meissner effect. 
Due to the action of the  Lorentz force on the radially outmoving electron it   acquires an azimuthal velocity given by \cite{copses}
\beq
v_\theta=-\frac{qr}{2mc} H
\eeq
which is identical to Eq. (37), the speed acquired by a charge in an orbit of radius $r$ when the applied magnetic field is increased from 
$0$ to its final value $H$. The reason is Faraday's law for Eq. (37), Lorentz force for Eq. (50).
When the radial motion is over a distance $r=2\lambda_L$ the azimuthal velocity acquired is
\beq
v_\theta=-\frac{q \lambda_L}{m c}H
\eeq
which is the same as the speed of the Cooper pairs  in the Meissner current Eq. (27). Note that it is important in this analysis that the magnetic field
imparting the azimuthal velocity to the electron in the expanding orbit is $H$ rather than $B$.

   \begin{figure}
 \resizebox{8.5cm}{!}{\includegraphics[width=6cm]{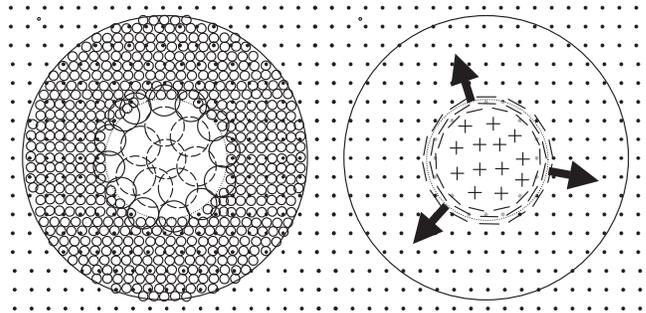}}
 \caption {Left panel:  In the superconducting region the carriers reside in orbits of radius $2\lambda_L$,
 in the normal region in orbits of radius $k_F^{-1}$.
 At the boundary of the normal-superconducting region (dotted circle)  the orbits expand from radius
$k_F^{-1}$ to radius $2\lambda_L$, causing magnetic field expulsion. 
Right panel: in the superconducting region the charge distribution is not homogeneous, there is an excess negative charge within
$\lambda_l$ of the surface of the superconducting region that spills over into the normal region, and an excess positive charge in its interior. }
 \label{figure1}
 \end{figure}

 Figure 7 shows schematically the expansion of a single orbit in a magnetic field pointing out of the paper. As the orbit expands the carrier's orbit cuts through magnetic
 field lines and in so doing acquires an azimuthal velocity, shown clockwise in Fig. 7 assuming the carrier has positive charge. When the radius reaches
 $2\lambda_L$, the azimuthal speed reaches the value Eq. (27). In turn, the motion of this charge generates a magnetic field in the direction opposite to the applied
 field, i.e. into the paper in Fig. 7. For a single carrier of course the magnitude of this counterfield is negligible.
 
 Now we consider the growth of the superconducting region in a cylinder. In the normal outer region the orbits are microscopic, in the interior
 superconducting region the orbits have radius $2\lambda_L$, and at the boundary between superconducting and normal regions
 the orbits are expanding. This is shown schematically in Figure 8 (left panel). The carrier in each expanding orbit reaches the final azimuthal velocity Eq. (27) when the
 radius of the orbit reaches its final value $2\lambda_L$.  Superposition of these
 motions results in a current being carried within a layer of thickness $\lambda_L$ from the boundary of the normal-superconducting region. 
 As the superconducting region expands, this boundary current expands with the boundary, and no net current remains in the interior region due to 
 cancellation of the internal motions. When the boundary
 reaches the boundary of the sample, the system reaches the superconducting state where all the magnetic field has been excluded except within a layer
 of depth $\lambda_L$ from the surface, and a Meissner current flows in this layer. 
 In this process  it is crucial that each carrier  is affected by the full external magnetic field as its orbit expands, rather than one that is partially compensated
  by the magnetic field created by the expanding
 orbits of other carriers, in order  that its azimuthal speed reaches the final value  Eq. (51).  Note in particular that within the enlarged orbits at the phase boundary in Fig. 8 (left panel) there are
 small normal orbits that will expand next, in a net magnetic field that is already weaker because of the field generated by the larger orbits enclosing them. However these
 smaller expanding orbits should not be affected by the magnetic field (in direction opposite to the applied field) created by the already enlarged orbits,
 because these carriers are all becoming part of the same $\Psi(\vec{r})$,
hence   they do not affect each other.  

      \begin{figure}
 \resizebox{8.5cm}{!}{\includegraphics[width=6cm]{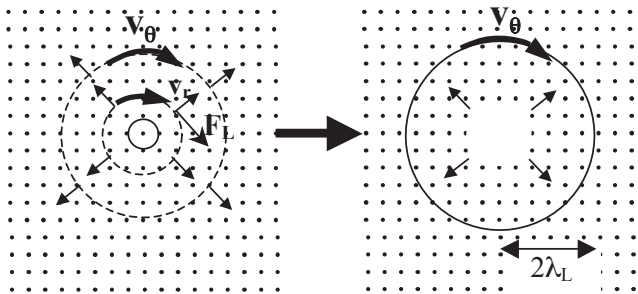}}
 \caption { A single orbit expanding in the magnetic field. As the orbit cuts through the field lines (left panel) 
 it acquires azimuthal velocity $v_\theta$ reaching the value Eq. (27) when the radius 
 reaches $2\lambda_L$. When the magnetic field lines move out ( right panel) they do not affect the
 azimuthal velocity because different parts of $\Psi(\vec{r})$ do not affect each other.   }
 \label{figure1}
 \end{figure}

 To clarify this point  further, we show in Fig. 9 the process of expansion of a single orbit.
 As the orbit cuts through the magnetic field lines the electron acquires azimuthal velocity
 $v_\theta$ due to the Lorentz force, which reaches the value Eq. (51) when the orbit
 reaches radius $2\lambda_L$. Then, the magnetic field lines move out cutting through
 the $2\lambda_L$ orbit because of the compensating magnetic field generated
by all the carriers in  $\Psi(\vec{r})$. This would slow down the azimuthal motion due to Faraday's law
{\it if this
 was a normal carrier}. Instead, because the carrier is part of $\Psi(\vec{r})$ its azimuthal
 speed is not affected   as shown in the right panel of Fig. 9.
 In reality, the processes shown on the left and right panels of Fig. 9 do not occur sequentially but simultaneously.

We conclude from this analysis that in the transition to superconductivity, in addition to the Cooper pairing predicted by the conventional theory, the individual carriers forming the
 Cooper pair expand their orbits from microscopic dimension ($k_F^{-1})$ to radius $2\lambda_L$, without being affected by the
 magnetic field generated by other carriers becoming part of the same macroscopic wavefunction, and that this physics resolves
 the puzzle of the Meissner effect.

In the right panel of Fig. 8 we show schematically the charge distribution. Because  the expansion of the orbits involves some outward charge motion, our theory predicts that associated with the $2\lambda_L$ orbits 
there is an excess negative charge near the boundary of the
superconducting region that spills over into the normal region \cite{chargeexp,electrospin}. Thus the outward motion of the phase boundary can be understood 
both as associated  with outward motion of negative charge into the normal region
and associated with enlargement of orbits from radius $k_F^{-1}$ to radius $2\lambda_L$.
The driving force for expansion of the orbits and associated   negative charge expulsion is lowering of quantum kinetic energy, which is what drives 
 superconductivity according to the theory of hole superconductivity \cite{kinenergy,dynhubscr}.

  \section{ $2\lambda_L$ orbits and the London moment}
 The London moment is the magnetic moment generated by a rotating superconductor \cite{londonbook}. The magnetic field that exists throughout the
 interior of a superconductor rotating with angular velocity $\vec{\omega}_0$ is
 \beq
 \vec{B}=-\frac{2m_e c}{e}\vec{\omega}_0
 \eeq
 The resulting magnetization for a cylinder rotating around its axis is
 \beq
 M=\frac{B}{4\pi}=-\frac{m_e c}{2\pi e}\omega_0
 \eeq
 Assuming as in Eq. (32) that each supercarrier contributes magnetic moment $\vec{\mu}$ to the magnetization yields
 \beq
 \mu=\frac{m_e c}{2\pi e n_s}\omega_0=\frac{2e}{c}\lambda_L^2\omega_0
 \eeq
 and the relation Eq. (34) between magnetic moment and angular momentum yields
 \beq
 \ell=m_e (2\lambda_L\omega_0)(2\lambda_L)
 \eeq
 which describes carriers in orbits of radius $2\lambda_L$ and tangential velocity $2\lambda_L\omega_0$.
 This provides additional support to the interpretation that carriers in the superconducting state reside in orbits of radius $2\lambda_L$.
 The fact that the London moment is parallel rather than antiparallel to the angular velocity demonstrates that the carriers
 forming the superfluid have negative charge \cite{lm}.
 
  \section{Spin Meissner effect}
  
  The Spin Meissner effect \cite{sm} is the spontaneous generation of a spin current within a 
  London penetration depth of the surface of a superconductor when a metal is cooled into
  the superconducting state, given by
  \beq 
  \vec{J}_\sigma=n_s\vec{v}_\sigma^0=-n_s\frac{\hbar}{4m_e\lambda_L}\vec{\sigma}\times \hat{n}
  \eeq
  where $\hat{n}$ is the outward-pointing  normal to the surface of the superconductor and 
  $\vec{\sigma}$ is parallel to the surface. The 
  magnitude of the mass transport associated with this current in each direction is   half the mass transport of  the critical current of the superconductor.
  When a magnetic field $\vec{B}$  is applied the spin current component with 
  $\vec{\sigma}$ parallel to $\vec{B}$ slows down  and the one with opposite spin
  direction speeds up, and when the magnetic field is such that the  slower spin current component is stopped
  the superconducting state is destroyed \cite{sm}.

  The negative charge expulsion predicted by our theory \cite{chargeexp} has as consequence that 
   an outward-pointing electric field
  exists in the interior of superconductors at sufficiently low temperatures. Because of this, a spin current
  originating in the Rashba spin-orbit interaction is expected. However, the magnitude
  of the spin current given by Eq. (56) is orders of magnitude larger than would be expected
  from the ordinary Rashba effect \cite{electrospin,holecore}.
  
  The spin-orbit interaction of an electron in the presence of an electric field $\vec{E}$ obtained from the Dirac
  Hamiltonian is
  \beq
  H_{s.o.}=-\frac{e\hbar}{4m_e^2c^2}\vec{\sigma}\cdot (\vec{E}\times\vec{p}) .
  \eeq 
  This can be represented by the Aharonov-Casher vector potential \cite{ac,ac2} $\vec{A}_\sigma$ in the single-particle Hamiltonian
  \bmath
  \beq
  H=\frac{1}{2m_e}(\vec{p}-\frac{e}{c}\vec{A}_\sigma)^2
  \eeq
  \beq
  \vec{A}_\sigma=\frac{\hbar}{4m_e c}\vec{\sigma}\times \vec{E} .
  \eeq
  \emath
  The term linear in $\vec{A}_\sigma$ from Eq. (58a) yields Eq. (57) (for an interpretation of the
  term quadratic in $A_\sigma$ see ref. \cite{slafes}). 
  Just like the ordinary vector potential $\vec{A}$ gives rise to a magnetic field $\vec{B}=\vec{\nabla}\times\vec{A}$,
  the spin-orbit vector potential $\vec{A}_\sigma$ gives rise to an effective magnetic field \cite{sm}
  \beq
  \vec{B}_\sigma=\vec{\nabla}\times\vec{A_\sigma}=\frac{\hbar}{4m_ec}(\vec{\nabla}\cdot\vec{E})\vec{\sigma}
  =\frac{\pi \hbar}{m_e c}\rho \vec{\sigma}
  \eeq
  that imparts an azimuthal velocity to the carriers in the expanding orbits, just as the ordinary magnetic
  field does\cite{copses}. Here, $\rho=\vec{\nabla}\cdot\vec{E}/4\pi$ is the charge density that gives rise to the
  electric field $\vec{E}$ with which the magnetic moment of the electron interacts.
  
  The question now is, what is this charge density $\rho$? The superfluid density $n_s$ has associated
  with it a charge density $qn_s$, with $q$ the charge of an individual carrier. 
  Experiments such as the London moment, the gyromagnetic effect and the Bernoulli effect indicate that the
  charge carriers in the superfluid are electrons, hence $q=e$ \cite{lm}. Again we have to assume that  the superconducting fluid described by $\Psi(\vec{r})$ does not screen itself
  and as a consequence the spin-orbit interaction affecting each superfluid carrier results from  the electric field generated by the $full$ compensating ionic
  charge density
  \beq
  \rho=|e| n_s 
  \eeq
  rather than from the slight net charge imbalance resulting from charge expulsion, which is smaller than Eq. (60) by a factor
  $v_\sigma^0/c$ \cite{electrospin}. 
  Replacing Eq. (60) in Eq. (59) and using Eq. (30) for $\lambda_L$ yields
  \beq
  \vec{B}_\sigma=\frac{\pi \hbar}{m_e c}|e|n_s \vec{\sigma}=-\frac{\hbar c}{4e\lambda_L^2}\vec{\sigma}  .
  \eeq
  When the orbit expands to radius $2\lambda_L$, the azimuthal speed acquired is
  \beq
v_\sigma^0=-\frac{e \lambda_L}{m_e c} {B}_\sigma=\frac{\hbar}{4m_e \lambda_L} .
  \eeq
  and just like for the Meissner effect, the internal motions cancel out and a spin current remains within
  $\lambda_L$ of the surface, given by Eq. (56). The direction of the spin current is as given in Eq. (56).
  The angular momentum of electrons in $2\lambda_L$ orbits with speed given by Eq. (62) is
  \beq
\ell=m_e v_\sigma ^0 \times (2\lambda_L)= \frac{\hbar}{2} ,
  \eeq
  the same as the intrisic electron angular momentum due to spin.
  
    The  condition Eq. (63) presumably has a topological origin and is   what determines that the orbits expand to radius $2\lambda_L$, which coincidentally is precisely what is needed to generate a magnetic
    field of just the right magnitude to cancel the external magnetic field and give rise to the Meissner effect. 
    The driving force for the orbit expansion is lowering of
    quantum kinetic energy  in the transition to the superconducting state \cite{slafes,kinenergy}. We regard 
    the fact that the result Eq. (63) results from this analysis to be compelling   evidence in favor of the 
    validity of this model for the description of real superconductors.

   \section{  macroscopic phase coherence}
   
   Macroscopic phase coherence is a hallmark of superconductivity \cite{tinkham,feyn,joseph,merc}. The BCS wavefunction exhibits macroscopic phase coherence, however it does not provide an
   intuitive picture of what macroscopic phase coherence means, nor how it is established in the transition from the normal to the superconducting state, nor how it is
   robustly maintained over macroscopic distances in the superconducting state, nor how the establishment of phase coherence is related to the Meissner effect. 
   
   Instead, the theory discussed here provides a unified explanation for how superconductors expel magnetic fields and how macroscopic
   phase coherence is established. Within our theory the superfluid wavefunction is composed of paired orbits of spin up and spin down electrons, each orbit of radius $2\lambda_L$, with distance
   between the centers of the orbits $\xi$, the superconducting coherence length \cite{sm}. We can think of the ``phase'' as a point in the electron's orbit  that is rotating with
   angular velocity $\omega = v_\sigma^0/(2\lambda_L)=|e|/(2m_ec)B_\sigma$.
   As the carriers condense into the superconducting state their orbits expand and overlap with each other, and this gives rise to phase coherence because
   overlapping orbits have to have the same phase to avoid collisions, as shown schematically in Fig. 10. We can easily understand
that  this phase coherence has to extend over the entire region occupied by the superfluid wavefunction $\Psi(\vec{r})$.

         \begin{figure}
 \resizebox{6.5cm}{!}{\includegraphics[width=6cm]{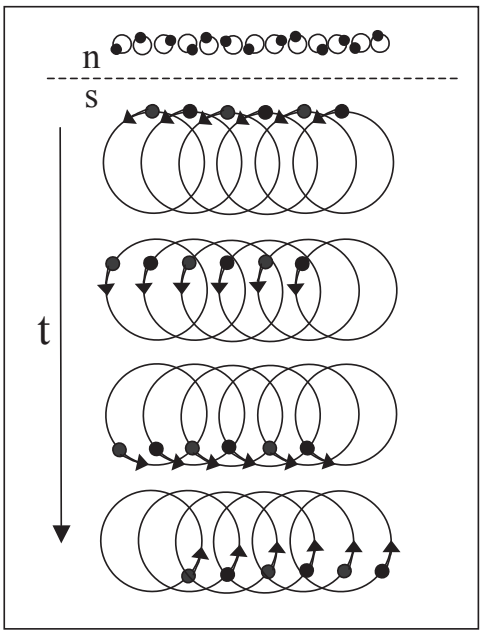}}
 \caption { The ``phase'' is depicted as a black circle on the orbit. In the normal (n) phase (small orbits of radius $k_F^{-1}$) there is no correlation between the phases of different orbits because they
 are non-overlapping. In the superconducting (s) phase (large orbits of radius $2\lambda_L$) the orbits overlap strongly  and the phases have to be the same in the different orbits to avoid collisions.
 As time (t) progresses the phases rotate together with angular velocity $\omega=(2\pi/\hbar)n_s\mu_B^2$.
 The overlap between the orbits enforces the long-range phase coherence.   }
 \label{figure1}
 \end{figure}

   \section{Summary and discussion}
  In this paper we have argued that the generally held view that the conventional theory of superconductivity describes, explains and
  predicts the Meissner effect is incorrect. The Meissner effect is the process by which a metal becoming superconducting
  expels the magnetic field from its interior.   It is generally a very non-trivial question whether a many-body system will reach its lowest energy state predicted by thermodynamics.
  For example a ferromagnet cooled below its critical temperature will not in general achieve an ordered state with macroscopic 
  magnetization but rather break up into domains. A liquid when cooled will often end up as a solid glass or a polycrystal rather than
  a macroscopically ordered single crystal.   In these processes the forces at play are well studied and understood. In contrast it is remarkable
  that it is generally assumed that this question is not relevant to superconductors. How superconductors achieve the state with the
  magnetic field expelled  is not regarded to be an open question in the field \cite{bcs50}.
The question has not even been posed, let alone been answered, in the vast literature dealing with the conventional theory of
  superconductivity.   
  We suggest that the question has not been posed nor answered because the conventional theory lacks essential
  physical ingredients that are necessary to answer it.

It has been argued that the explanation (or lack thereof) of the Meissner effect is in essence the same as that of flux
quantization in superconducting rings \cite{nikulov,leggett}. We argue that this is not quite so, even though the phenomena are
certainly  closely related. Flux quantization involves changes in a fraction of the flux quantum
\beq
\phi_0=\frac{hc}{2e}=\frac{\pi}{\alpha} e
\eeq
with $\alpha=e^2/(\hbar c)\sim 1/137$ the fine structure constant. Eq. (64) is a microscopic quantity ($\sim 430$ electron charges).
How superconducting rings manage to adjust their current flow in order to respect flux quantization is certainly a
fundamental question, as discussed extensively  by A.V. Nikulov \cite{nikulov}, for which we don't have a dynamical explanation. However, expulsion of a $200$ Gauss magnetic field
from a sample of cross section $1cm^2$ is a $much$ $bigger$ question: it involves getting rid of some $10^9$ flux quanta! This is a macroscopic
phenomenon for which it is reasonable to expect an explanation that is consistent with the macroscopic laws of physics.
The fact that we don't have a dynamical explanation of flux quantization is not in our view a valid reason to argue that
a dynamical explanation of the Meissner effect is not required from the 60-year old theory that claims to explain conventional
superconductivity \cite{leggett}.

 It is very suggestive that the process of expulsion of magnetic field and transition to the superconducting state of a macroscopic sample
 occurs through $radial$  $expansion$ of small superconducting regions (Fig. 1(b) or 1(c)), rather than uniformly as depicted in Fig. 1(a). 
 We argue that this known experimental fact 
 gives a vivid image of the underlying physics, and have provided  here  two 
 complementarly closely related physical explanations:
 (i) $expansion$ of a perfectly conducting fluid (Fig. 2(b)) leads to expulsion of magnetic field, as is well known in plasma physics
 (Alfven's theorem \cite{plasma,plasma2}), and (ii) $expansion$ of electronic orbits leads to  increased Larmor diamagnetism (Eq. (40)), as is well known
 in atomic and solid state physics \cite{am}. In both processes the 
 $dynamical$ explanation for the magnetic field expulsion is the magnetic Lorentz force \cite{lorentz} acting on radially outgoing charge.
The driving force for a radial $expansion$ is naturally a $pressure$, in our interpretation {\it quantum pressure} driven by
 reduction of quantum kinetic energy \cite{emf,kinenergy,helium}. The postulated outflow of superconducting carriers and
 backflow of normal carriers (Fig. 4) strongly resembles processes known to exist in $^4He$, where superfluid thrusts from colder to warmer
 regions and normal fluid backflows from warmer to colder regions (fountain effect) \cite{londonbook2}. In superfluid $^4He$ the driving force for the transition
 is known to be lowering of quantum kinetic energy \cite{helium}. Finally,    within our model   the radial expansion of the phase boundary also gives a macroscopic
 image of the microscopic {\it atomic orbital expansion} that gives rise to superconductivity   as described by
 the dynamic Hubbard model \cite{dynhubscr}.
 
 In contrast, within conventional BCS theory there is no radial charge flow associated with the radial expansion of the phase boundary, hence the driving force for the azimuthal current cannot be the
 magnetic Lorentz force and remains unidentified. The driving force for the radial expansion of the phase boundary is termed
 `Meissner pressure' by F. London \cite{londonbook} but is not given a physical interpretation, and as a consequence the expansion of the phase boundary   
bears no relationship to  the underlying microscopic 
 physics believed  to be responsible for superconductivity within BCS \cite{bcs}, 
 namely the Fr\"ohlich electron-phonon interaction and Cooper pairing. 
  It could be said that   conventional BCS theory describes the `expansion' of phase coherence as the superconducting nucleus expands. Then,
 BCS theorists have to explain how  the expansion of phase coherence causes an azimuthal force to act on charge carriers. Perhaps this is a new
force of nature that has not yet been identified \cite{weinberg2} and will make BCS become part of the `reductionist frontier' \cite{weinberg}.
 In our model instead, no new forces are needed and  the development of phase coherence is explained by the expansion of the orbits to become strongly overlapping
 (Fig. 10), which in turn is directly linked to the azimuthal force acting on charge carriers (Fig. 9).

We argue that since the conventional theory does not describe charge expulsion it cannot describe the Meissner effect. More generally, 
a theory of superconductivity that can explain the Meissner effect  by expulsion of charge has to know the difference between positive and
negative charge, just as superconductors do \cite{lm}. We argue that this rules out all theories of superconductivity that are
electron-hole symmetric, as most theories including the conventional theory are. 
Furthermore, since the expulsion of charge carries along an increase in potential (Coulomb) energy, we argue that this
 rules out any theory of superconductivity where the condensation energy is potential rather than kinetic which is the
 case of most theories including the conventional theory. Within our theory, electron-hole asymmetry and kinetic energy lowering
 are inextricably linked \cite{dynhubscr}.
 
 We have shown that the magnetization that the superconductor develops to cancel the applied magnetic field originates
 in the orbital magnetic moments of electrons residing in mesoscopic orbits of radius $2\lambda_L$. 
 The charge expulsion discussed in the previous paragraphs originates in the orbit enlargement from microscopic radius $k_F^{-1}$ to
 radius $2\lambda_L$ in the transition to superconductivity \cite{electrospin}. In the absence of
 an applied magnetic field, electrons in these orbits give rise to a Spin Meissner effect \cite{sm}, the existence of
 a macroscopic spin current within a London penetration depth of the surface of superconductors. The dynamical
 generation of the spin current and the expulsion of external magnetic fields occur through the same process,
 the expansion of the electron orbit from microscopic radius to the mesoscopic radius $2\lambda_L$ in the presence of   external
 magnetic field and internal electric field from the background ionic charge distribution. 
 The fact that the speed of electrons in these orbits gives rise to angular momentum of value precisely
 $\hbar/2$ \cite{sm} we regard as compelling evidence that the theory applies to nature.
 
 Finally we have pointed out that to understand both the Meissner and the Spin Meissner effects it is necessary to assume
 that different parts of the macroscopic superfluid wavefunction $\Psi(\vec{r})$ do not influence each other \cite{schr},
 i.e. the magnetic field generated by electrons in overlapping orbits does not affect the magnetic field sensed by a given electron,
 and the ionic background electric field giving rise to the spin-Meissner current is not screened by the charge of electrons
 in overlapping orbits.
 This is consistent with the fact that the macroscopic wavefunction of the superconductor $\Psi(\vec{r})$ is in many ways similar
 to the wavefunction  $\Psi(\vec{r})$ of a single electron \cite{scully}.

 The process of negative charge expulsion and existence of mesoscopic orbits described here gives rise   to a small charge inhomogeneity over the
 entire macroscopic sample with the region within $\lambda_L$ of  the surface  having a small excess negative charge. 
 The resulting macroscopic equilibrium electrodynamics equations giving the spatial distribution and quantitative values of
 the charge density, electric field and spin current in the ground state of superconductors are given
 in other publications \cite{chargeexp,electrospin}. 
 A valid microscopic theory of superconductivity will have to be
 consistent with these macroscopic and mesoscopic properties.

\acknowledgements
The author is grateful to D.J. Scalapino, A. J. Leggett and N. Goldenfeld for discussions on the conventional understanding of the Meissner effect.

\end{document}